\documentclass[12pt,preprint]{aastex}

\begin{document}


\title{The Relationship between the Sudden Change of the Lorentz Force and the Magnitude of Associated Flares}


\author{Shuo Wang, Chang Liu and Haimin Wang}
\affil{Space Weather Research Laboratory, New Jersey Institute of Technology,\\University
Heights, Newark, NJ 07102-1982, USA}
\email{shuo.wang@njit.edu}



\begin{abstract}

The rapid and irreversible change of photospheric magnetic fields associated with flares has been confirmed by many recent studies. These studies showed that the photospheric magnetic fields respond to coronal field restructuring and turn to a more horizontal state near the magnetic polarity inversion line (PIL) after eruptions. Recent theoretical work has shown that the change in the Lorentz force associated with a magnetic eruption will lead to such a field configuration at the photosphere. The Helioseismic Magnetic Imager has been providing unprecedented full-disk vector magnetograms  covering the rising phase of the solar cycle 24. In this study, we analyze 18 flares in four active regions, with GOES X-ray class ranging from C4.7 to X5.4. We find that there are permanent and rapid changes of magnetic field around the flaring PIL, the most notable of which is the increase of the transverse magnetic field. The changes of fields integrated over the area and the derived change of Lorentz force both show a strong correlation with flare magnitude. It is the first time that such magnetic field changes have been observed even for C-class flares. Furthermore, for seven events with associated CMEs, we use an estimate of the impulse provided by the Lorentz force, plus the observed CME velocity, to estimate the CME mass. We find that if the time scale of the back reaction is short, i.e., in the order of 10 s,  the derived values of CME mass ($\sim 10^{15}$g) generally agree with those reported in literature.

\end{abstract}





\section{Introduction}

Solar flares have been understood as an energy release process due to magnetic reconnections in the solar corona \citep{kop76}. The magnetic fields in the solar corona are anchored in the dense photosphere. Historically, the photospheric magnetic fields were assumed to be unaffected by flares on short time scales because of high mass density there. Of course their long-term evolution is well known to play an important role in  storing the energy and triggering the flares.

\cite{wang92} and \cite{wang94} were the first to show observational evidence of flare-related rapid/irrevisible change of photospheric magnetic fields  based on ground-based vector magnetograms.  The most striking but controversial finding at that time was the  increase of magnetic shear along the magnetic polarity inversion line (PIL). Using line-of-sight magnetograms of SOHO/MDI, \cite{kos01} found that some irreversible variations of magnetic field in the lower solar atmosphere happened very rapidly in the vicinity of PILs at the beginning of the flare of 2000 July 14. \cite{wan02} analyzed the observed photospheric magnetic flux evolution across 6 X-class flares, and found significant permanent changes associated with all the events. After surveying 15 X-class flares, \cite{sud05} concluded that the change in the line-of-sight (LOS) magnetic field always occurs during X-class flares.  \cite{wan06} noticed that when an active region is away from the disk center, the reconnected low-lying fields would cause an apparent increase of the flux in the polarity toward the limb and a decrease in the polarity closer to the disk center.

Until the launch of SDO, these studies were very limited due to the paucity of continuous/consistent high-quality vector magnetogram series. With a nearly continuous coverage over the entire solar disk, vector magnetograms are being obtained from the Helioseismic and Magnetic Imager \citep[HMI;][]{schou11} on board the Solar Dynamics Observatory (SDO),  making possible extensive studies that achieve a fundamental physical understanding of the observations. A number of recent papers using HMI data have all pointed to the same conclusion that photospheric magnetic fields turn more horizontal immediately after flares and that magnetic shear increases at surface but relaxes in the corona \citep{wan12,sun12,liuc12}. For example, \cite{wan12} found a
rapid (in about 30 minutes) and irreversible enhancement in the horizontal magnetic field at the flaring magnetic PIL by a magnitude of $\sim$ 30\% associated with the X2.2 flare on 2011 February 15. \cite{petrie12} has analyzed the magnetic field evolution and Lorentz forces in the X2.2 flare on 2011 February 15, and also found a large Lorentz force change coinciding with the eruption.

From the theoretical side, \cite{hud08} quantitatively assessed the back reaction on the photosphere and solar interior with the coronal field evolution required to release flare energy, and predicted that the magnetic field should become more horizontal after flares. \cite{wan10} were first to link this idea to observed field changes.  They provided observational evidence of the increase of transverse field at the PIL when vector magnetograms were available. When only the LOS field measurement was available,  they found that if the source active region is not located at the disk center, the measured apparent LOS field changes are consistent with the picture of \cite{hud08}, i.e., fields turn more horizontal across the PIL. They used the same concept which we mentioned before: due to the projection effect, there is an apparent increase of the flux in the polarity toward the limb and a decrease for the polarity closer to the disk center. More recently, \cite{fisher12} and \cite{hud12} further developed analytic modeling, by separately considering Lorentz forces acting on the upper solar atmosphere and the solar interior. The upward momentum of the erupting plasma can be estimated by equating the change in the upward momentum with the Lorentz force impulse acting on the outer solar atmosphere. The authors also argued that the back reaction on the solar interior may be responsible for the sudden sunspot motion on the photosphere and the excitation of seismic waves in the interior.

It is noted that the previous studies were mainly focused on large flares such as X-class or upper M-class events. HMI has been obtaining seeing-free, high-resolution data since 2010 April. In this Letter, we target our study on the magnetic field change associated with flares in a broad range of magnitudes, including C-class events.  We also attempt to find the possible relationship among flare magnitude, field changes, and momentum involved in the eruptions.

In Section 2, we will describe observations and data processing, and will show two examples of case studies.  The statistical results between flare magnitude and field changes will be presented in Section 3, in which we will also discuss a practical method to estimate the CME mass. Section 4 will give the summary and discussion.

\section{Observations and Data Processing}


HMI and the Atmospheric Imaging Assembly \citep[AIA;][]{lemen11} on board SDO provide full-disk, multi-wavelength observations in high spatial and temporal resolution. The LOS magnetic field observation with $\sim$0.5$''$ pixel scale and a 45s cadence has recorded all flares on the solar disk since April 2010. The noise level of the LOS field measurement is 10 G. HMI also provides full-disk vector magnetic field measurement with a separate  system. However, the measurement has larger uncertainty due to the difficulty in the Stokes inversion. It is noted that HMI team has put significant effort to improve the inversion code. Using an average of 12-minute data, the accuracy of the transverse field measurement is in the order of 100 G \citep{hoeksema12} as derived from Stokes Q and U. We are using the latest release of the processed data from the HMI data archive. The HMI vector fields are derived using the Very Fast Inversion of the Stokes Vector (VFISV) algorithm \citep{borrero11} based on the Milne Eddington approximation. The 180$^{\circ}$ azimuthal ambiguity is resolved with the minimum energy method \citep{metcalf94,leka09}. Four active regions that produced X-class flares in 2011 and 2012 have been analyzed in this study. They are NOAA regions 11158, 11166, 11283, and 11429.




To demonstrate the procedure of data handling, let us first describe the analysis of the largest flare in our sample: the X5.4 flare in AR 11429.
AR 11429 was located in the northeast of the solar disk when the X5.4 flare peaked at 00:24 UT on 2012 March 07. The data cube covers the entire active region with projection effect corrected.

For this and all the events under study,  we first scrutinize the movie of the horizontal fields covering the flare. Rapid/irreversible enhancement of the horizontal field is always clearly shown across the flaring PIL that can be identified with the help of the corresponding AIA images (Figure 1(d)).  To pinpoint the location of the horizontal field change, we construct a difference image by subtracting a postflare horizontal field image (Figure 1(b)) from a preflare horizontal field image (Figure 1(a)). We then use the contour with a level of 120 G (slightly above the confidence level of the measured transverse field) as the boundary of the region of interest (ROI) for further quantitative analysis. In Figure 1(c), the ROI covers part of the flaring PIL. We then plot the mean field change in the ROI as a function of time.  As shown in Figure 1(e), the horizontal magnetic field within the ROI increases by $\sim$40\% from $\sim$650 G to $\sim$920 G in $\sim$30 minutes. This is consistent with all the previous studies that showed a step-like change of the fields. The observed field change is clearly above the fluctuation (indicated by the error bars) by more than an order of magnitude. In addition, we did not detect any transient changes of the fields due to flare emissions as described by \cite{patte81} and \cite{qiu03}.

Next, we analyze the related Lorentz force change as formulated by \cite{hud08} and \cite{fisher12}. Here, we focus on the Lorentz force change in the volume below the photosphere using the equation in Fisher et al. (2012):

\begin{equation}
\delta F_{r}=\frac{1}{8\pi}\int_{A_{ph}}dA(\delta B_{r}^{2}-\delta B_{h}^{2}),
\end{equation}

\noindent where $B_{r}$ is the vertical field while $B_{h}$ is the horizontal field. We note that the radial field shows no rapid irreversible change in any of the 18 events (see Figures 1(e) and 2(e) as examples). Therefore, we omitted the term \begin{math} \delta B_{r}^{2} \end{math} to minimize the effect of longer term evolution. The summation of \begin{math} \delta F_{r}\end{math} in the whole ROI gives the value of the integrated Lorentz force change. The total change of Lorentz force in the volume below the photosphere during this flare is \begin{math} 1.1 \times 10^{23}\end{math} dyne, comparable with previous studies.

The same data analysis procedure is applied to all the 18 events in four active regions. The result of the C4.8 event on 2011 Feburary 15 is shown in Figure 2. This event did not occur in the main PIL that produced the large X2.2 flare on the same day. However, the stepwise increase of the horizontal field is clearly demonstrated.

\section{Statistical Results}

After studying the horizontal field movies for all observed events in these four ARs, we find that 18 flares (listed in Table 1) obviously show a rapid/irreversibe change in the horizontal fields.  The magnitude of flares ranges from GOES-class C4.7 to X5.4.  In these active regions, all the M2.0 and above flares have the detectable changes. Three out of five M1.0--M1.9 flares and five out of 17 C4.0--C9.9 flares also demonstrate a similar pattern of field change.  As described in the previous section, the ROI is defined using a threshold of 120 G based on the difference image between the horizontal fields right before and after flares. We then calculate the integrated horizontal magnetic field change and the downward Lorentz force change. For each event, we find that (1) the ROI is spatially related to the flare kernels pinned down with AIA data, covering the flaring PIL; and (2) the evolution of magnetic field and the downward Lorentz-force change both show variations in a stepwise fashion. Based on this statistical study, we find significant correlations between the peak GOES X-ray flux and both the integrated horizontal field change and the total downward Lorentz force change. The results are shown in Figure 3.

In panels (a), (b), (c), and (d) of Figure 3, we plot the integrated horizontal field enhancement, the total change of the downward Lorentz force, ROI size, and the mean horizontal field change respectively as  functions of the peak soft X-ray flux of flares. It is clear that the amount of field change is correlated well with flare magnitude. The linear fit on a log-log scale gives a cross correlation coefficient of around 0.8 and a slope (corresponding power index) of around 0.5 for the first three parameters (Figures 3 (a)--(c)). The last parameter (Figure 3(d)) is not sensitive to the flare magnitude. We use the unit of magnetic flux Mx to describe the integrated  horizontal field. In reality, it is not magnetic flux as the horizontal field is not normal to the surface. We note a previous study of \cite{chen07}, in which the darkening of sunspot intensity at the flaring PIL line is also related to the flare magnitude, consistent with the picture of fields turn to horizontal. However, in that study, the authors were not able to analyze the magnetic structure change.

This is the first time that the rapid/irreversible field changes are found to be associated with C-class flares. Of course, we need to be careful about the confidence level of the detected changes. In Figures 1(e) and 2(e), to demonstrate the fluctuation before and after the flare, we plot the $3\sigma_{\rm pre}$ ($3\sigma_{\rm pos}$) as error bars, where $\sigma_{\rm pre}$ ($\sigma_{\rm pos}$) is derived from the linear fit of the temporal evolution of the horizontal field in the preflare (postflare) state. As shown in Figure 2(e), the rapid change of the horizontal field even for the C4.8 flare is significant compared to variations seen in the long-term evolution.


Our main motivation of this study is to estimate the change of the Lorentz force acting on either the outer solar atmosphere or on the solar interior. One important application is to evaluate the upward momentum associated with magnetic eruptions. The upward impulse exerted on the outer solar atmosphere is suggested
to account for the CME momentum \citep{fisher12}. Therefore the estimated CME mass is:

\begin{equation} M_{CME} \simeq \frac{1}{2} \frac{\delta F_{r} \delta t}{v}, \end{equation}

\noindent where here, $\delta F_r$ is the change of the Lorentz force acting on
the outer solar atmosphere (with the same magnitude but opposite sign as $\delta F_{r}$ in Eq.~1), ${v}$  is the CME speed available from the LASCO CME catalog \footnote{http://cdaw.gsfc.nasa.gov/CME\_list}, and $\delta t$ is the change over time of the field. In using this expression, we've made the assumption that the entire mass moves with the same velocity, a gross over-simplification, and we have also ignored the work done against gravity. It does, nevertheless, provide an indpendent estimate for the CME mass. Due to the 12 minute cadence of HMI data, it is difficult to evaluate the critical parameter $\delta t$ for the CME mass estimate. There is evidence that the back-reaction is impulsive \citep{donea12,sud05,petrie10}. We therefore use the typical time scale of hard X-ray spike time, i.e., around 10 s, as a rough estimate of $\delta t$ during the flare impulsive phase. We also assume that the initial CME velocity is zero. The estimated masses of the seven CME events are shown in Table 2, and are consistent with the typical value in the previous studies \citep{vourl10,carley12}. Please note that among the 18 events, these seven are the only ones that have identified CMEs in the LASCO catalog. Unfortunately, the mass estimates of these CMEs are not yet available from LASCO white-light intensity analysis to be compared with our derived values.

\section{Summary and Discussion}

Taking advantage of the newly released HMI vector magnetograms in flare-productive active regions,  we are able to analyze changes of vector magnetic fields associated with 18 flares. This is the first time that such changes are found for small flares down to the GOES C class. The results listed in Tables 1 and 2 agree with previous studies in the following aspects:
\begin{enumerate}
  \item All events exhibit a step-like increase of the horizontal magnetic field after flares, with an order of magnitude of \begin{math} 10^{20}\end{math} Mx after integrating over the ROI.
  \item  The changes are co-temporal with the flare initiation, and the change-over time is about three time bins of the HMI data, i.e., 36 minutes. However, we believe that the reaction time for the field change could be much shorter than this.
\end{enumerate}

From  the statistical studies of the 18 events, we also note the following:
\begin{enumerate}
  \item The permanent magnetic field change is always co-spatial with the PIL connecting the two primary flare kernels.
  \item Significant linear relationships between the peak GOES X-ray flux and all the following parameters are found: the size of the affected area, the integrated horizontal field change, and the total downward Lorentz force change.
\end{enumerate}

The above findings clearly support the idea of back reaction of surface magnetic fields to the eruption in the corona, as proposed by \cite{hud08} and \cite{fisher12}.  The fields are observed to change from a more vertical to a more horizontal configuration. The downward change of Lorentz force reflects such a topological change in magnetic fields. In the photospheric layers, in static equilibrium before and after eruptive events, there should be a balance between the Lorentz force, gas pressure gradients, and gravity. The Lorentz force difference between the post-flare and pre-flare states is the signature of an unbalanced Lorentz force in the solar atmosphere, occurring during the time of the eruption, in which Lorentz forces are balanced primarily by the inertial force of the accelerating plasma.

If the above physics is correct, then the upward CME momentum can be estimated based on the derived impulse associated with the Lorentz force change. We can then estimate the CME mass.  However,  as we already mentioned, an uncertain parameter in the equation is the reaction time associated with the field change. We prefer to use a short time (10 s based on the hard X-ray observation), as the change is observed to occur in a time scale close to the temporal resolution of the HMI data. If a longer time is used, the estimated CME mass will be much larger than the established values in literature. It is  easier to  estimate mass of CMEs for the close-to-limb events based on the white-light image intensity such as that measured by LASCO coronagraph. We are providing an independent method to estimate the CME mass based on the change of the photospheric magnetic field.  This is particularly useful for events closer to the disk center. Our positive correlation between the change of Lorentz force and the peak soft X-ray flux also agrees with the study of \cite{zhangj04} and \cite{zhangj06}, in which they found that the CME speed is associated with the soft X-ray flux.



\acknowledgments
The authors thank the SDO/MHI and AIA teams for obtaining outstanding vector magnetograms and EUV data. We also thank the SOHO/LASCO team for the CME catalog, which is generated and maintained at the CDAW Data Center by NASA and the Catholic University of America in cooperation with the Naval Research Laboratory. We thank the referee for valuable comments that helped us to improve this Letter. The research is supported by NSF grants, AGS-0745744, AGS-0839216, and AGS-0849453, and NASA grants NNX08AJ23G and NNX11AC05G.

\clearpage

\begin{figure}
\epsscale{1.0}
\plotone{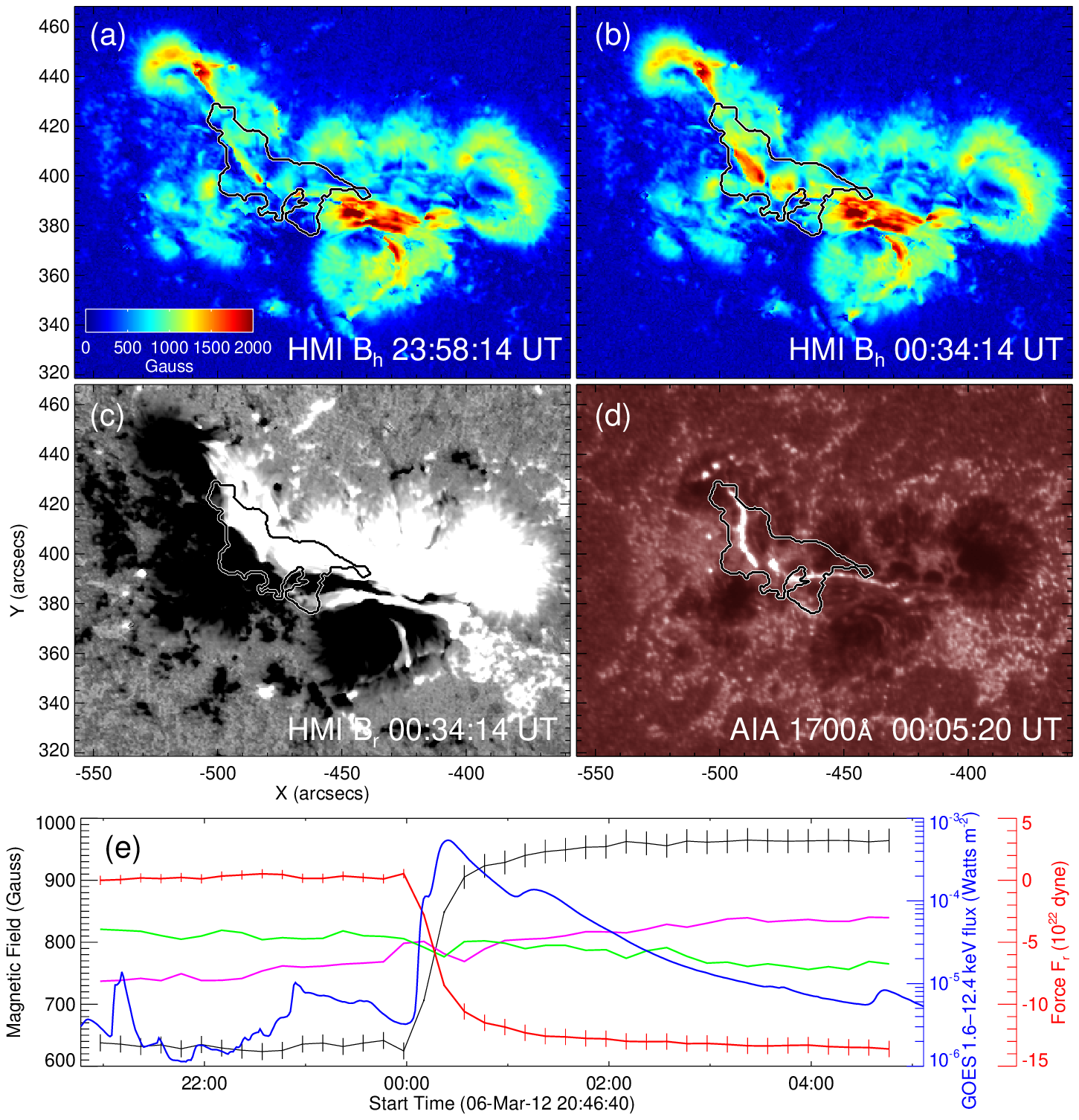}
\caption{X5.4 flare on 2012 March 07. Panels (a) and (b) show the preflare and postflare horizontal magnetic field maps. Panel (c) is the radial magnetic field map. Panel (d) is an AIA 1700~\AA\ map. The ROI is overplotted with the white-bordered black contour. In the panel (e), the black and red curves with vertical error bars are the temporal evolution of the mean horizontal magnetic field and radial Lorentz force within the ROI respectively, in comparison with the GOES light curve in 1--8~\AA\ (blue curve). The vertical error bars indicate a 3$\sigma$ level of the fluctuation in the pre- and postflare states. Purple and green curves represent positive and negative mean radial magnetic fields, respectively, which do not show obvious step wise changes. \label{figx54} }
\end{figure}

\begin{figure}
\epsscale{1.0}
\plotone{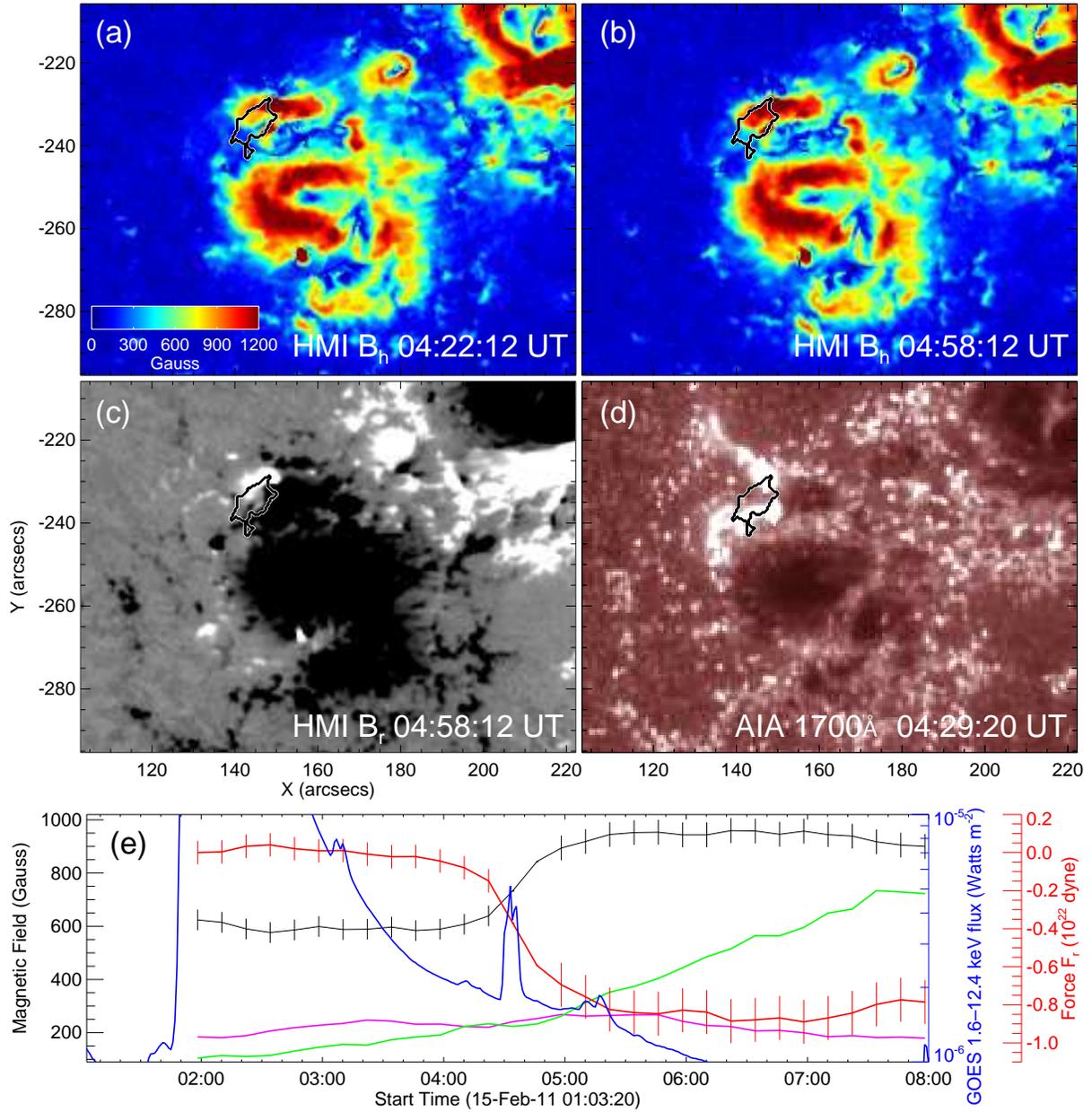}
\caption{Same as Figure 1, but for the C4.8 flare on 2011 February 15. This small flare occurred in a different PIL that produced the X2.2 event around 02 UT. \label{figc48} }
\end{figure}

\begin{figure}
\plotone{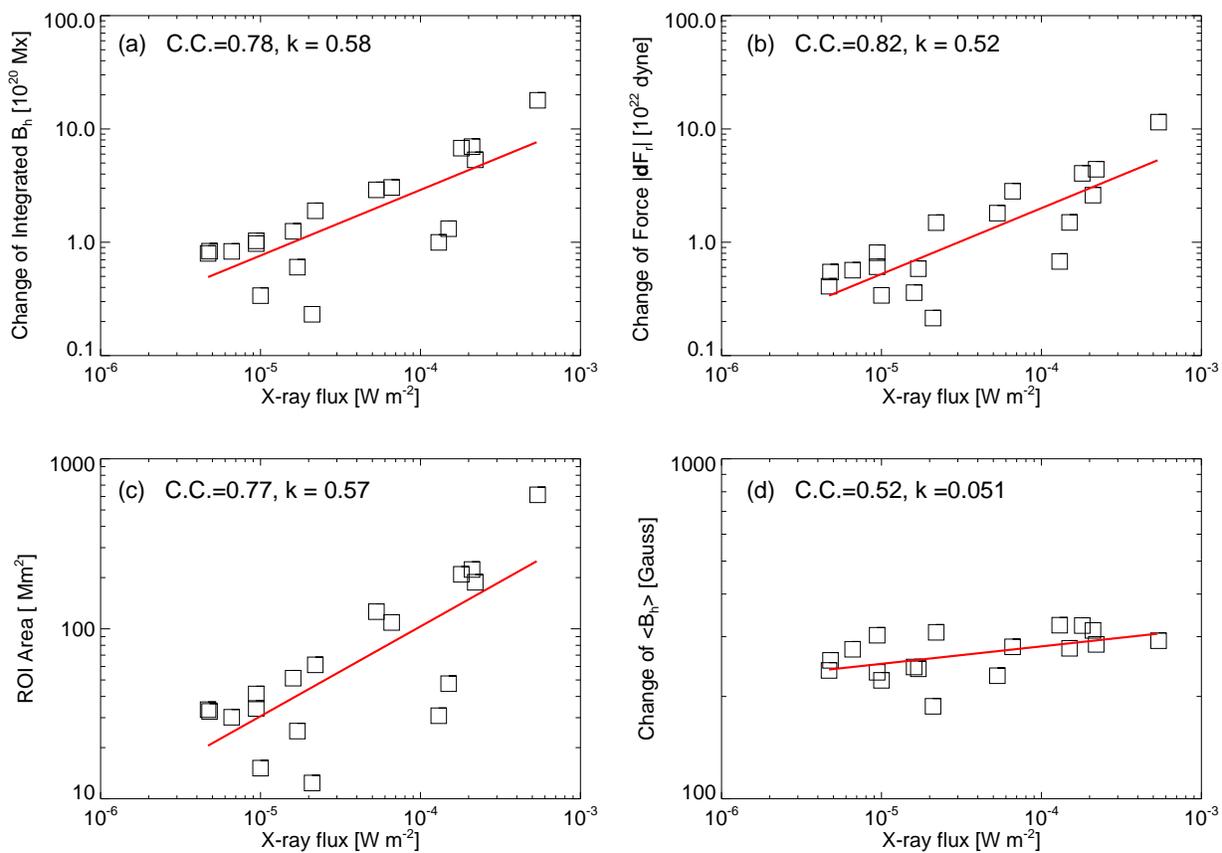}
\caption{Scatter plots of the peak GOES X-ray flux in 1--8~\AA\ vs. various parameters. The red lines show the least-squares linear fit to the data points. The correlation coefficient (C.C.) and slope (k, corresponding to power index in linear-linear plot) are shown in each panel. \label{figfitting}}
\end{figure}

\begin{table}
\caption{Events with Rapid Change of Horizontal Photospheric Magnetic Fields} 
\centering 
\small
\begin{tabular}{c c c c c} 
\hline\hline 
GOES 1--8~\AA\ &  NOAA  &  GOES  &  Integrated $\delta B_h$  &  Total $\delta F_r$ \\ [0.5ex] 
Peak (UT) & AR & Class & (\begin{math}10^{20}\end{math} Mx) & (\begin{math}10^{22}\end{math} dyne) \\ [0.5ex]
\hline 
2011 Feb 13 13:56 & 11158 & C4.7 & 0.80 & 0.41 \\
2011 Feb 13 17:38 & 11158 & M6.6 & 3.0 & 2.8 \\
2011 Feb 14 12:53 & 11158 & C9.4 & 1.0 & 0.81 \\
2011 Feb 14 17:26 & 11158 & M2.2 & 1.9 & 1.5 \\
2011 Feb 14 19:30 & 11158 & C6.6 & 0.83 & 0.57 \\
2011 Feb 15 01:56 & 11158 & X2.2 & 5.3 & 4.4 \\
2011 Feb 15 04:32 & 11158 & C4.8 & 0.83 & 0.55 \\
2011 Feb 16 14:25 & 11158 & M1.6 & 1.2 & 0.36 \\
2011 Mar 09 14:02 & 11166 & M1.7 & 0.97 & 0.61 \\
2011 Mar 09 22:12 & 11166 & C9.4 & 0.60 & 0.59 \\
2011 Mar 09 23:23 & 11166 & X1.5 & 1.3 & 1.5 \\
2011 Sep 06 01:50 & 11283 & M5.3 & 2.9 & 1.8 \\
2011 Sep 06 22:20 & 11283 & X2.1 & 7.0 & 2.6 \\
2011 Sep 07 22:38 & 11283 & X1.8 & 6.8 & 4.1 \\
2012 Mar 06 07:55 & 11429 & M1.0 & 0.34 & 0.34 \\
2012 Mar 06 12:41 & 11429 & M2.1 & 0.23 & 0.21 \\
2012 Mar 07 00:24 & 11429 & X5.4 & 17 & 11 \\
2012 Mar 07 01:14 & 11429 & M1.3 & 1.0 & 0.68 \\ [1ex] 
\hline 
\end{tabular}
\label{table:bhchange} 
\end{table}

\begin{table}[ht]
\caption{Events with CME} 
\centering 
\small
\begin{tabular}{c c c c c c} 
\hline\hline 
GOES 1--8~\AA\  & NOAA  & GOES  & CME  & CME Speed & CME Mass \\
Peak (UT) & AR & Class & Time (UT) & (km s\begin{math}^{-1}\end{math}) &
(\begin{math}10^{15}\end{math} g) \\ [0.5ex] 
\hline 
2011 Feb 13 17:38 & 11158 & M6.6 & 18:36 & 373 & 3.8 \\
2011 Feb 14 17:26 & 11158 & M2.2 & 18:24 & 326 & 2.3 \\
2011 Feb 15 01:56 & 11158 & X2.2 & 02:24 & 669 & 3.3 \\
2011 Mar 09 23:23 & 11166 & X1.5 & 23:05 & 332 & 2.3 \\
2011 Sep 06 01:50 & 11283 & M5.3 & 02:24 & 782 & 1.2 \\
2011 Sep 06 22:20 & 11283 & X2.1 & 23:05 & 575 & 2.3 \\
2011 Sep 07 22:38 & 11283 & X1.8 & 23:05 & 792 & 2.6 \\
[1ex] 
\hline 
\end{tabular}
\tablecomments{Information of the CME time (the first C2 appearance time) and the CME
speed are from LASCO CME catalog. The masses were computed assuming $\delta t = 10$s (see Eq.(2)).}
\label{table:cmemass} 
\end{table}





\end{document}